\documentclass[aps, prapplied, reprint, amsmath, amssymb, longbibliography, superscriptaddress]{revtex4-2}

\usepackage[T1]{fontenc}
\usepackage{palatino}
\usepackage{soul}
\usepackage{wasysym}
\soulregister\cite7
\soulregister\ref7
\soulregister\pageref7
\soulregister\ce7
\usepackage{graphicx}
\usepackage[svgnames]{xcolor}
\usepackage{amsmath, bm}
\usepackage{fancyref}
\usepackage[colorlinks,
            linkcolor=Blue,
            urlcolor=Blue,
            citecolor=DarkGreen]{hyperref}
\usepackage[utf8]{inputenc}
\usepackage{textcomp}
\usepackage[version=3]{mhchem}
\usepackage{miller}
\usepackage{lipsum}

\newcommand{\orcidicon}[1]{\href{https://orcid.org/#1}{\includegraphics[height=\fontcharht\font`\B]{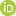}}}

\graphicspath{{Figures/}}

\usepackage{pdfpages}
\usepackage{pgffor}
\usepackage{etoolbox}
\makeatletter
\patchcmd{\@outputpage@head}{\@ifx{\LS@rot\@undefined}{}{\LS@rot}}{}{}{}
\makeatother

\begin{document}

\title[]{Oxide Membranes from Bulk Micro-Machining of SrTiO3 substrates}

\author{Nicola~\surname{Manca}\,\orcidicon{0000-0002-7768-2500}}
\email{nicola.manca@spin.cnr.it}
\affiliation{CNR-SPIN, C.so F.\,M.~Perrone, 24, 16152 Genova, Italy}
\affiliation{\textsc{RAISE} Ecosystem, Genova, Italy}
\author{Alejandro~E.~\surname{Plaza}\,\orcidicon{0000-0002-2538-7585}}
\affiliation{CNR-SPIN, C.so F.\,M.~Perrone, 24, 16152 Genova, Italy}
\affiliation{\textsc{RAISE} Ecosystem, Genova, Italy}
\author{Leon\'elio~\surname{Cichetto}~Jr\,\orcidicon{0000-0002-5894-8852}}
\affiliation{CNR-SPIN, C.so F.\,M.~Perrone, 24, 16152 Genova, Italy}
\author{Warner~J.~\surname{Venstra}\,\orcidicon{0000-0003-1731-050X}}
\affiliation{Quantified Air BV, Langegracht 70, 2312NV Leiden, The Netherlands}
\author{Cristina~\surname{Bernini}\,\orcidicon{0000-0003-4807-0924}}
\affiliation{CNR-SPIN, C.so F.\,M.~Perrone, 24, 16152 Genova, Italy}
\affiliation{\textsc{RAISE} Ecosystem, Genova, Italy}
\author{Daniele~\surname{Marré}\,\orcidicon{0000-0002-6230-761X}}
\affiliation{Dipartimento di Fisica, Università degli Studi di Genova, 16146 Genova, Italy}
\affiliation{CNR-SPIN, C.so F.\,M.~Perrone, 24, 16152 Genova, Italy}
\author{Luca~\surname{Pellegrino}\,\orcidicon{0000-0003-2051-4837}}
\affiliation{CNR-SPIN, C.so F.\,M.~Perrone, 24, 16152 Genova, Italy}
\affiliation{\textsc{RAISE} Ecosystem, Genova, Italy}


\begin{abstract}
Suspended micro-structures made of epitaxial complex oxides rely on surface micro-machining processes based on sacrificial layers. These processes prevent to physically access the microstructures from both sides, as substantial part of the substrate is not removed. In this work, we develop a bulk micromachining protocol for a commonly used substrate employed in oxide thin films deposition. We realize suspended oxide thin film devices by fabricating pass-through holes across \ce{SrTiO3(100)} or \ce{SrTiO3(110)} substrates. Careful calibration of anisotropic etching rates allows controlling the final geometry of the aperture in the substrate in a predictable way. As demonstrators of possible device geometries, we present clamped membranes and trampolines realized from deposited thin films of \ce{(La{,}Sr)MnO3}, a conductive magnetic oxide, and a suspended trampoline resonator carved from the \ce{SrTiO3} substrate itself. Reported protocols can be readily applied to a broad variety of other complex oxides so to extend the use of membranes technology beyond those of commercially available silicon compounds.
\end{abstract}

\maketitle

\section{Introduction}

Recently, there has been an increasing interest in the fabrication of suspended or
freestanding structures made of complex oxides~\cite{Chiabrera2022}.
The release processes include the use of various sacrificial layers \cite{Bakaul2016, Paskiewicz2016, Shen2017, Higuchi2018, Li2019}, even water-soluble\cite{Lu2016}, the fabrication of superconducting oxide interfaces by spalling \cite{Sambri2020, Erlandsen2022}, the integration of complex oxides with silicon suspended structures \cite{Niu2009, Reiner2010, Baek2013, Liu2014, Torres2016, Dong2018, Spreitzer2021}, or the fabrication of suspended oxide structures by etching their growth substrate \cite{Pellegrino2009, Deneke2011}. These methods enabled new kinds of experiments, taking advantage from the lower thermal dissipation of suspended thin films \cite{Manca2015, Manca2015a}, the possibility to apply large tensile strain \cite{Hong2020, Xu2020}, or from the measurement of the mechanical modes of oxide-based resonators \cite{Davidovikj2020, Manca2023}. Applications where suspended oxides are particularly promising include pressure and gas sensors \cite{Manca2019a, Lee2022b}, mechanical resonators \cite{Manca2017b, Manca2022}, bolometers \cite{Mechin1997, Nascimento2021}, nano-actuators \cite{Manca2020, Manca2021}, piezoelectric devices \cite{Takahashi2020, Lee2022a}, or photo-strictive systems \cite{Ganguly2024}. However, all these examples are based on the paradigm of surface micro-machining, while actual bulk micro-machining, i.e. the possibility to shape the growth substrate considering its whole thickness, is still missing for complex oxides. Bulk micro-machining on silicon allowed the integration of proof masses in accelerometers, gyroscopes, or seismic sensors, the fabrication of large membranes, the 3D integration by wafer bonding, and physical access from backside to enable optical readout~\cite{Corigliano2018, Romero2020}. 

Here, we present a fabrication protocol to realize suspended devices from epitaxial oxides thin films with pass-through holes on their backside. This process is based on the wet chemical etching of the growth substrate across its full thickness. We discuss the general aspects of bulk micro-machining for \ce{SrTiO3} substrates having (001) and (110) cut-planes and analyze their anisotropic etching rates in detail. We then show three examples of suspended oxide devices realized with this protocol: trampoline resonators with controlled backside aperture width, sealed membranes, and trampoline resonators carved from the \ce{SrTiO3} substrate itself. 

\section{Results and Discussion}

The fabrication protocol presented in this work relies on the selective wet chemical etching of \ce{SrTiO3} (STO). Its key element is the use of double-polished substrates having epitaxial thin films of a different oxide grown on both sides. These films are employed as hard masks during the substrate wet etching or even as device layers to realize fully suspended structures. The use of oxide films as hard masks is motivated by their perfect adhesion provided by the epitaxial growth. This is not the case for polymers deposited by spin-coating or metal layers grown by thermal evaporation, which typically detach during the long acid baths required by sample processing. The basic condition for the mask material is to be resistant to HF that we employ to etch the STO substrates. Several complex oxides are suitable for this task, but here we choose \ce{(La{,}Sr)MnO3} (LSMO). The reason is that LSMO is a magnetic compound already proposed for several device applications \cite{Zhang2021, Nascimento2021, Vera2023, Enger2023}. It also grows tensile strained on STO, resulting in flat structures with no wrinkles and relatively high mechanical quality factor due to the dissipation-dilution mechanism \cite{Engelsen2024, Manca2025}. LSMO hard masks are deposited on both sides of the polished substrate by pulsed laser deposition as discussed in our previous works~\cite{Plaza2021}. X-ray diffraction analysis showing $\phi$-scans and $\theta$--2$\theta$ measurements of the LSMO films are reported in the Supporting Information Sec.~I. They confirm epitaxial growth mode of LSMO and show that no significant difference can be identified in the film strain condition after immersion in HF. A critical point of this method is the handling of the substrate itself. Since the STO is clamped to a stainless-steel holder that would contaminate the backside during the growth, we place a second STO substrate in between the sample and the holder. This second layer introduces a slightly increased thermal resistance from the sample holder to the growth substrate. This must be considered when setting up the deposition conditions. Patterning of both layers is realized by standard UV lithography as discussed in the ``Experimental Section''. The top and bottom patterns were aligned to each other using the optical microscope integrated with the mask aligner, taking advantage from the transparency of the STO substrate.

\subsection{Etching rates calibration}

\begin{figure}[]
	\includegraphics[width=\linewidth]{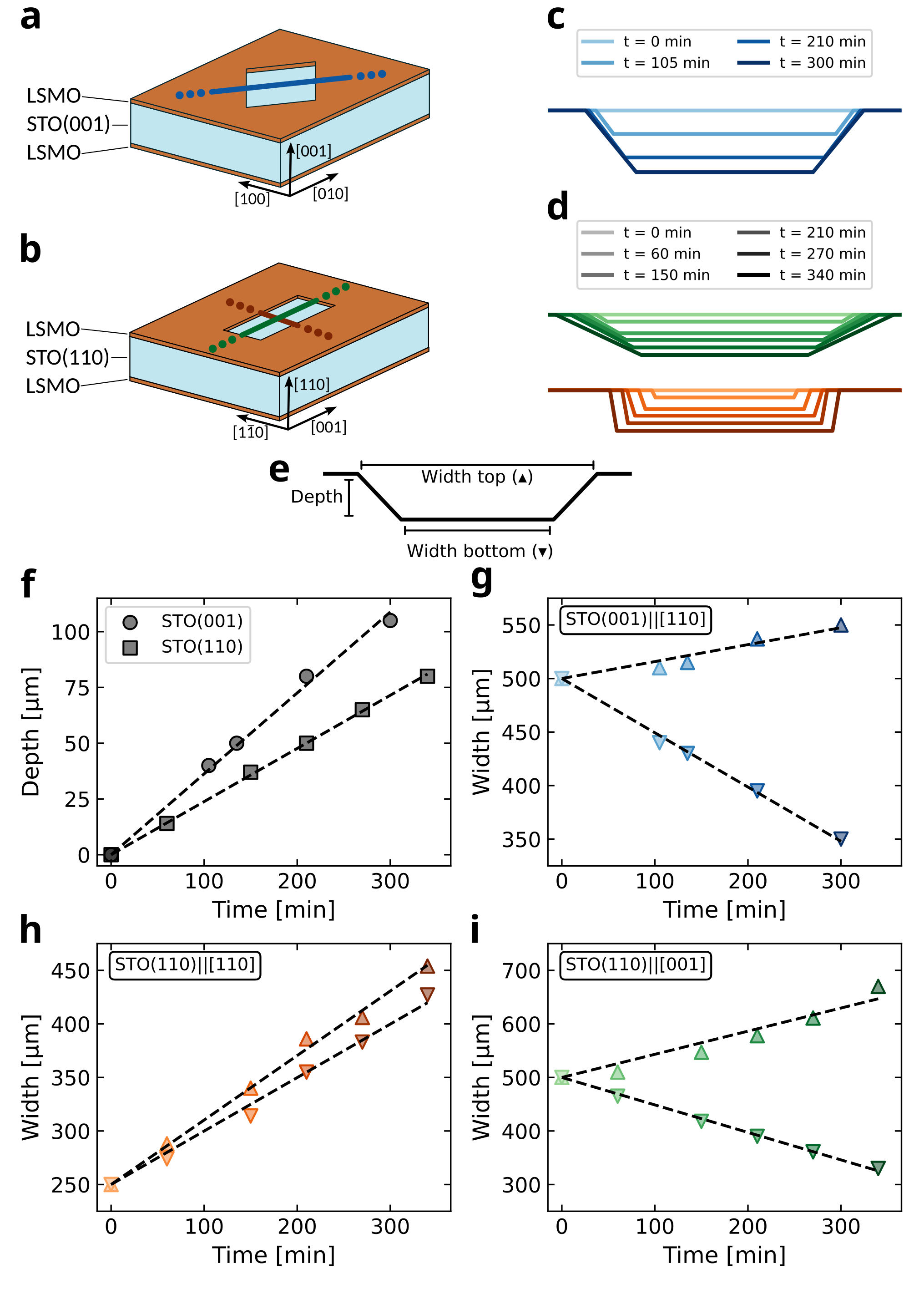}
	\caption{\label{fig:rate}
		Calibration of the etching rates of \ce{SrTiO3}(001) and (110) substrates.
		(a, b) Schematic of the two samples used to calibrate the rates. STO is light blue,  LSMO is brown, and the cut-lines indicate the directions along which the (c) and (d) profiles were measured.
		(c, d) Time evolution of the aperture profiles along the cross-sections in (a) and (b).
        (e) Diagram of the geometric parameters monitored during the etching process.
        (f)--(i) Time evolution of the depth ($\circ$, $\square$), the top width ($\vartriangle$), and the bottom width ($\triangledown$) of the apertures geometry as shown in (e). Cut-plane and cross-section direction are indicated by labels in each panel. The colors of the data-point in (g)--(i) correspond to those in (c) and (d).
	}
\end{figure}

The realization of a desired device geometry is only possible by careful designing the hard mask and the device layer, with precise knowledge of the etching evolution along different lattice directions. Previous studies indicate that the out-of-plane etching rate of STO in 5\,\% HF water solution at 30\,{\textcelsius} is 13.5\,{\textmu}m/h for (001) substrates and 9\,{\textmu}m/h for (110) substrates \cite{Plaza2021}. This means that etching through a 100\,\textmu{}m-thick substrate requires more than 7\,h for STO(001) and more than 11\,h for STO(110). Moreover, STO dissolution in HF results in by-products that are not soluble and progressively clog the STO surface, with a lowering of the dissolution rate and the accumulation of residues that are difficult to remove afterwards. To overcome these issues, we employed an acid solution consisting of a mixture of 5\,\% HF and 5\,\% \ce{H3PO4} in purified water and kept it at 40\,{\textcelsius}. The higher temperature speeds up the etching rate, while we observed that \ce{H3PO4} reduces the amount of the deposited etching by-products (such as \ce{SrF2} \cite{Plaza2021}, which is not soluble in water) and thus improves the cleanliness of the final structures.

To calibrate the etching rates in these experimental conditions, we prepared two samples by depositing a 100\,nm-thick LSMO layer on both sides of a STO(001) and a STO(110) substrate. For each sample, we realized a window in the LSMO top layer, exposing the STO substrate, and then periodically monitored the size of the resulting aperture while soaking the samples in the acid bath. The geometries of the apertures are shown in the schematic illustrations of Fig.~\ref{fig:rate}a and b, which were chosen based on a previous study reported in Ref.~\cite{Plaza2021}. For the STO(001) case, the hard mask is a 500$\times$500\,{\textmu}m$^2$ square window rotated by 45{\textdegree} with respect to the sample's edges, i.e. with the sides parallel to the $\langle$110$\rangle$ directions. This is because, for this cut-plane, the mask edges aligned along the $\langle$110$\rangle$ lattice directions preserve their orientation during the etching process, resulting in a good reproducibility in fabrication. For the STO(110) case, instead, the in-plane etching is anisotropic, and we thus employed a 250$\times$500\,{\textmu}m$^2$ rectangular window whose short/long edges are aligned to the fast/slow etching directions. The size of the apertures was large enough to be sure that they would maintain their shape throughout the whole etching time, about 7\,h, while still being fully visible within the field of view of the microscope employed for measuring their size. 

\begin{table}[]
	\begin{ruledtabular}
		\begin{tabular}{c c l c c}
			Cut-plane & Direction & Position & $r_{p/\perp}$ [{\textmu}m/min] \\
			\hline
			(001) & [001]     & depth  & +0.36 \\
			& $\langle$110$\rangle$     & top    & +0.16 \\
			& $\langle$110$\rangle$     & bottom & -0.50 \\
			\hline
			(110) & [110]     & depth  & +0.24 \\
			& $\langle$001$\rangle$     & top    & +0.43 \\
			& $\langle$001$\rangle$     & bottom & -0.51 \\  
			& $\langle$1\={1}0$\rangle$ & top    & +0.60 \\
			& $\langle$1\={1}0$\rangle$ & bottom & +0.50 \\  
		\end{tabular}
	\end{ruledtabular}
	\caption{\label{tab:rates} Etching rates of \ce{SrTiO3} in HF 5\,\% + \ce{H3PO4} 5\,\% in \ce{H2O} at 40\,\textcelsius. Different etching ``directions'' and ``positions'' are referred to the illustrations of Fig.~\ref{fig:rate}a, b, and e. Numerical values corresponds to the slopes of the linear fits (black dashed lines) reported in Fig.~\ref{fig:rate}f--h.}
\end{table}

The geometric parameters of the apertures formed in the LSMO window during the wet etching process are measured along the cut-lines indicated in Fig.~\ref{fig:rate}a and b, which are aligned with the following directions: STO(001)$\parallel$[110] (blue), STO(110)$\parallel$[001] (green), and STO(110)$\parallel$[1\={1}0] (red). These colors are used consistently across Fig.~\ref{fig:rate}. Details of the measurement procedure are discussed in the ``Experimental Section'' and Supporting Information Sec.~II. The reconstructed profiles are reported in Fig.~\ref{fig:rate}c and d. 
They provide a direct comparison between the aperture profiles at different times and for each lattice direction.
The shape of an aperture through STO at the end of the etching process, i.e.\ when the full thickness of the substrate is dissolved, can be predicted by knowing the time evolution of the depth, the top width, and the bottom width of the aperture, as schematically illustrated in Fig.~\ref{fig:rate}e. In this context, the main parameter is the out-of-plane etching rate ($r_\perp$), determining the time ($\Delta t$) required to cover the thickness of a substrate ($d$). The top/bottom width of the aperture after $\Delta t$ can be calculated as
\begin{equation}
  \label{eq:width}
  w_{\mathrm{f}} = w_{\mathrm{i}} + r_p \Delta t ~; \Delta t = d / r_{\perp}
\end{equation}
where $r_p$ is the in-plane etching rate, and $w_{\mathrm{i/f}}$ the initial/final width value. The time evolution of the aperture depth for both the substrates is reported in Fig.~\ref{fig:rate}f, while the time dependence of the top ($\vartriangle$) and bottom ($\triangledown$) width of the apertures is reported in Fig.~\ref{fig:rate}g--i, where each panel shows one of the lattice directions corresponding to the cut-lines of Fig.~\ref{fig:rate}a, b.
Etching rates are obtained from a linear fit of these data and are reported in Table~\ref{tab:rates}, where positive/negative values along the in-plane directions correspond to a widening/narrowing with respect to the initial LSMO mask width.
The error associated to the measurements of the geometrical parameters of the apertures is quite low, of about 2\,\%, resulting in an uncertainty of the etching rates of 0.01\,{\textmu m}/min. However, these etching rates are subject to much wider variations across different samples due to extrinsic effects that are difficult to control, such as temperature fluctuations, stirring speed of the acid bath, accumulation of etching by-products, or density of crystal defects. Because of this, their calibration in the specific experimental conditions may be necessary.

It is noteworthy that the sidewalls of the aperture are not bound to specific lattice planes. Their expected slope ($\theta_\mathrm{SW}$) along the measured directions can be calculated from the linear etching rates reported in Table\,\ref{tab:rates}
\begin{equation}
	\label{eq:angle}
	\theta_\mathrm{SW} = \arctan\left({\frac{r\mathrm{{_p}^{top}} - r\mathrm{{_p}^{bottom}}}{2 r_{\perp}}}\right)
\end{equation}

However, these sidewalls' angles are size-dependent because related to the microscopic evolution of the etching front in STO. Etch pits formed on the STO(001) surface and having small lateral size (less than 5\,{\textmu m}) show sidewalls parallels to \{111\} STO lattice planes.
Over time, new etch pits form along these exposed surfaces, locally enhancing the etching rate. The top width of the aperture widens as the etching in the in-plane directions gets promoted by defects, and, at the same time, the out-of-plane etching makes the aperture deeper.
As a result the slope of the sidewalls changes over time departing from the ideal \{111\} plane. A more detailed discussion is reported in the Supporting Information sec.~III.

\begin{table}[]
	\begin{ruledtabular}
		\begin{tabular}{c c c}
			Cut-plane & Direction & Sidewall angle [deg] \\
			\hline
			(001) & $\langle$110$\rangle$		& 42.5    \\
			(110) & $\langle$001$\rangle$		& 63.0 \\
			(110) & $\langle$1\={1}0$\rangle$	& 11.8 \\
		\end{tabular}
	\end{ruledtabular}
	\caption{\label{tab:angles} Sidewalls angle calculated from the etching rates of Table~\ref{tab:rates} and using Eq.~\ref{eq:angle}.}
\end{table}

\subsection{Fabrication protocol of through-hole devices}

\begin{figure}[]
	\includegraphics[width=\linewidth]{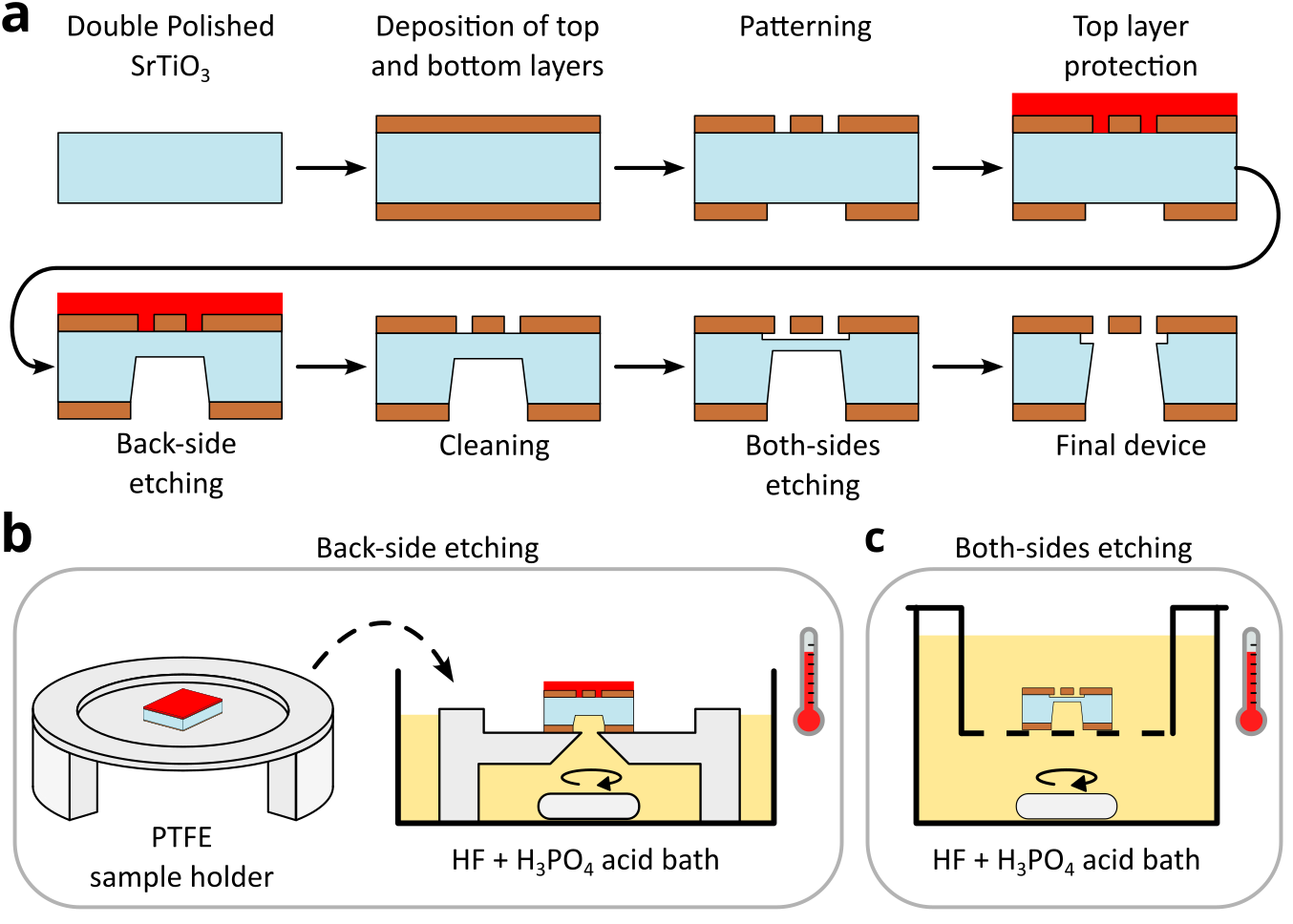}
	\caption{\label{fig:proc}
		Fabrication protocol of through-hole suspended devices on single crystal \ce{SrTiO3} substrates.
		(a) Main steps of the substrate micro-machining process.
		Schematic illustrations of (b) the setup for back-side wet etching and (c) the final wet etching step by full soaking the sample.
	}
\end{figure}

Figure~\ref{fig:proc}a shows a schematic step-by-step illustration of the fabrication protocol employed to realize oxide devices suspended on through-hole STO substrates.
The first step of the process is to thin down the substrate below the patterned device. This is required because a full etching of the substrate starting from the top would result in excessive under-etching. Back-side etching is performed in a custom Poly-tetrafluoroethylene (PTFE) sample holder as is depicted in Fig.~\ref{fig:proc}b. It has a hole at its center with diameter of $\sim$3\,mm to let the acid solution reach the sample. This hole is smaller than the typical substrate size (5\,mm) to prevent spillovers to the edges. The holder is placed in a PTFE cup containing the acid solution, which is kept at 40\,{\textcelsius} in bain-marie and agitated by a magnetic stirrer at about 200\,RPM. Since during this process both sides of the sample are already patterned, the unprotected STO on the top surface could be affected by HF vapors. To prevent this issue, we deposited a protection layer of SPR-220-4.5 photoresist on the top surface of the sample. Other compounds can be also employed, as long as they  have good adhesion, are resistant to HF, and allow for easy removal. When the thickness of the substrate in the aperture is below 20\,{\textmu}m, the sample is washed in distilled water. The protection layer is then removed from the top surface by ultrasonic bath in acetone and then ethanol. The final etching step consists in soaking the sample in the acid solution (5\,\% HF + 5\,\% \ce{H3PO4} in distilled water). To do so, we employ a transparent polypropylene holder (see Fig.~\ref{fig:proc}c) with the same temperature and agitation conditions as in the back-side etching step. This final step was implemented to suspend the device on the top layer before the opening of the hole across the substrate, which could otherwise affect the fabrication yield. Once the through-hole is completed, the sample is dried in a \ce{CO2} critical point drying system. 

\subsection{Trampoline resonators with backside aperture}

\begin{figure}[b]
	\includegraphics[width=\linewidth]{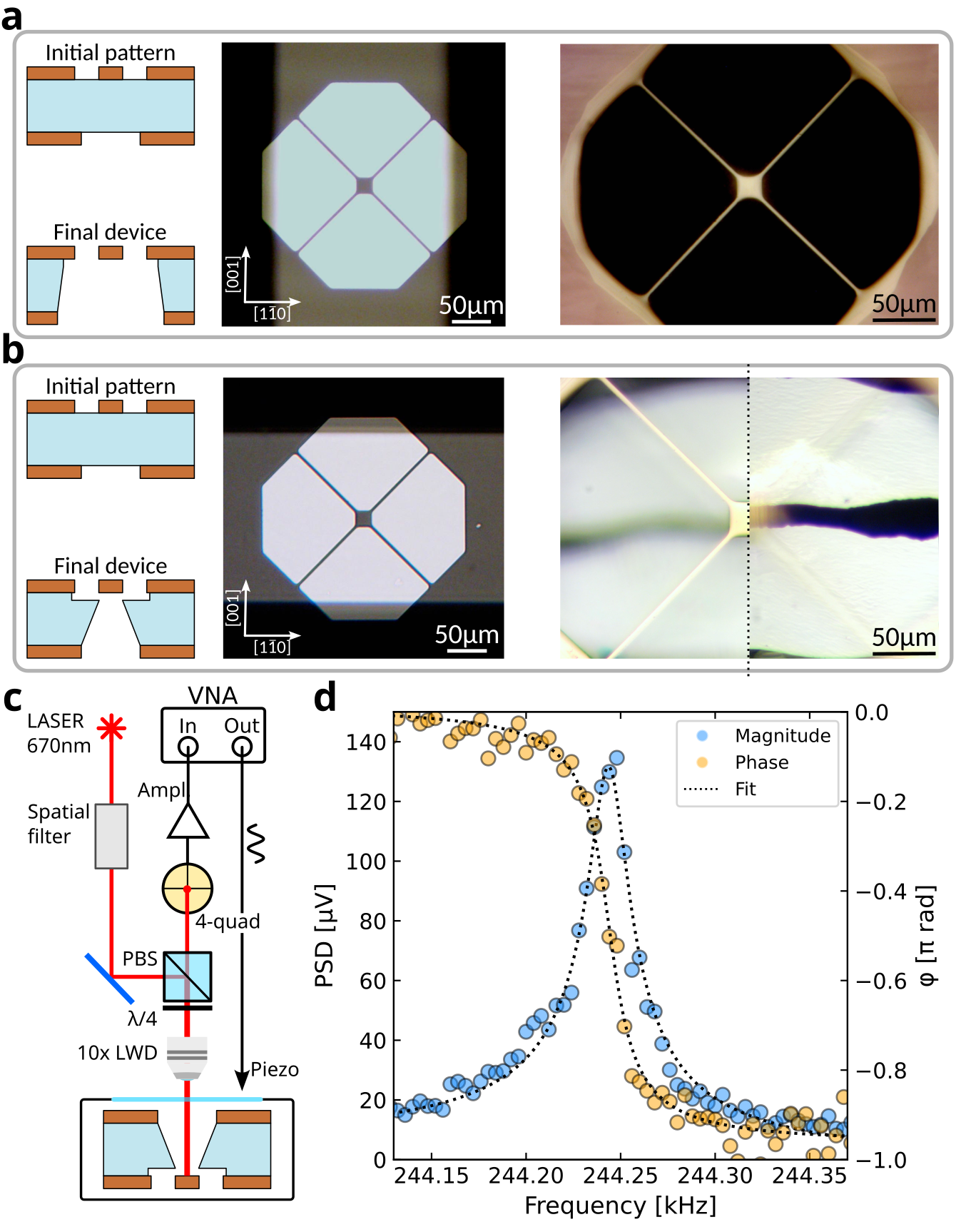}
	\caption{\label{fig:trench}
		Free-hanging trampoline with controlled aperture width in \ce{SrTiO3(110)}.
		(a, b) Schematic of the process, hard mask alignment and final device  with short edge parallel to the (a) [1\={1}0] or (b) [001] direction. The device in (b) is shown for the focal planes of the trampoline (left) and the trench (right), located about 20\,{\textmu}m below.  
		(c) Schematic of the optical setup used to measure the motion of the trampoline in (b). Details are reported in the ``Experimental Section''.
		(d) Spectrum of the first resonance mode (circles). Lorentzian fit (dashed line) indicates a $Q$-factor of 11700.
	}
\end{figure}

The fabrication protocol reported in Fig.~\ref{fig:proc} is employed to realize suspended trampoline resonators of LSMO. We took advantage of the etching anisotropy of STO(110) to control the width of the aperture in the substrate just below the suspended structure. We employed two identical samples having 100\,nm-thick LSMO films deposited on the top and bottom surfaces. The top layers are patterned as an array of trampolines having central square pad of 20$\times$20\,{\textmu}m$^2$ and 100\,{\textmu}m-long tethers. The bottom layers are patterned as a 200\,{\textmu}m-wide slot, but in one case the slot width is aligned along the [1\={1}0] direction while in the other along the [100] direction.
Figure~\ref{fig:trench}a and b show optical micrographs of the patterned LSMO layers (in transmitted light) and the final devices, together with schematic illustrations of the corresponding transversal sections. In Fig.~\ref{fig:trench}a the trench at the end of the process is wider and with almost vertical walls, in agreement with the positive etching rate in Table~\ref{tab:rates}, making the trampoline fully exposed from the bottom side. In Fig.~\ref{fig:trench}b, instead, the aperture is just a narrow trench in the substrate, whose final width was calibrated to correspond to the size of the trampoline's pad. To do so, accordingly with Eq. (\ref{eq:width}), we first removed 70\,{\textmu}m of substrate by back-side etching, as in Fig.~\ref{fig:proc}b), which corresponded to a trench width of about 55\,{\textmu}m. We then removed the last 40\,{\textmu}m by both-sides etching (Fig.~\ref{fig:proc}c). This further narrows the trench for only half of the etched thickness, making it about 20	\,{\textmu}m-wide.
As also specified in Ref.\cite{Plaza2021}, the etching of the substrate under the LSMO mask and pattern produces undercut, as visible in the optical image of Fig.~\ref{fig:trench}a (yellow regions). The width of the undercut increases with the immersion time and depends on the direction for each substrate orientation. This means that appropriate choice of pattern orientation must be done when designing a suspended structure. We note that this kind of device geometry allows us to employ the residual substrate as a hard mask for the deposition of additional layers, such as for example metallic or dielectric mirrors, on just the backside of the trampoline pad. This could be helpful to preserve the mechanical properties of the resonator by minimizing the added mass and the stress change of the tethers.

Despite the narrow aperture, it is possible to probe the motion of the LSMO trampoline from the backside. To do so, we employ the setup schematically reported in Fig.~\ref{fig:trench}c and described in the ``Experimental Section''. The response when the device is driven around the first mechanical resonance is shown in Fig.~\ref{fig:trench}d, together with the fit of magnitude (blue) and phase (orange) to a damped driven harmonic oscillator (black dashed line). The $Q$-factor is about 11700, which is in line with what measured on LSMO trampolines obtained by surface micro-machining and also in the ballpark of previous reports on LSMO micro-bridge resonators \cite{Manca2022, Manca2025}, indicating that the fabrication process did not critically affect the mechanical properties.

\subsection{Sealed oxide membrane}

\begin{figure}[b]
	\includegraphics[width=\linewidth]{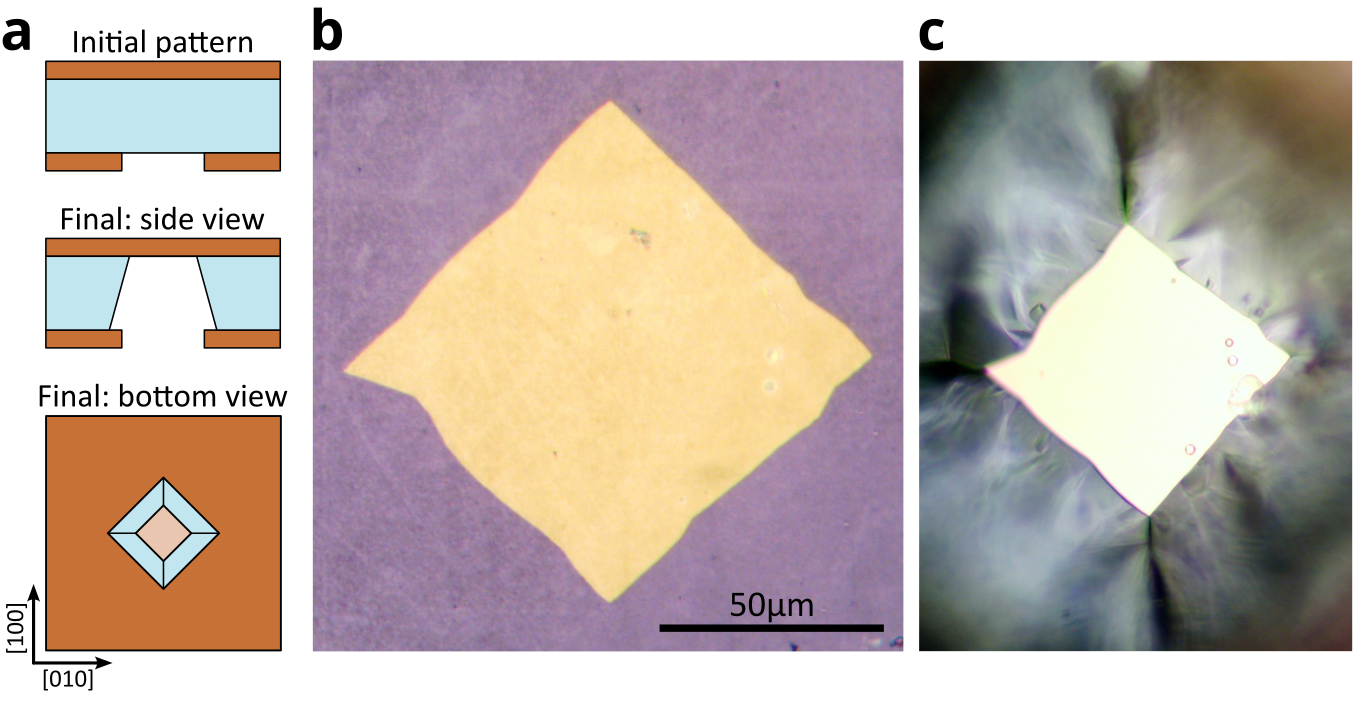}
	\caption{\label{fig:memb}
		LSMO clamped membrane.
		(a) Schematic illustration of the initial and final sample geometry.
		(b, c) Optical micrograph of the LSMO membrane seen from (b) top and (c) bottom in reflected light (not to-scale).
	}
\end{figure}

A simple modification of the bulk micromachining protocol described above allows for the realization of another kind of device: clamped membranes. In this case, the protocol reported in Fig.~\ref{fig:memb} is slightly modified and only the bottom layer undergoes a lithographic process. The schematic illustration of the initial and the final geometry of the sample is shown in Fig.~\ref{fig:memb}a. It was made of 100\,nm-thick LSMO films deposited on both sides of a 105\,{\textmu}m-thick STO(001) substrate. The bottom layer was patterned as a square rotated by 45{\textdegree}, similarly to Fig.~\ref{fig:rate}a, with a side length of 200\,{\textmu}m. The sample was etched from one side, as show in Fig.~\ref{fig:proc}b, with the top layer protected by SPR-220-4.5 photoresist, which was removed in acetone at the end of the process. The total etching time was 270\,min, resulting in a square membrane with 80\,{\textmu}m-long edges parallel to those of the initial hard mask. Fig.~\ref{fig:memb}b and c are optical micrographs of the membrane at the end of the fabrication process observed in reflected light from top and bottom, respectively. The edges of the membrane are not perfectly straight. This is because they are the result of the backside etching of the STO(001) substrate. As previously discussed, the etching process is driven by the formation and coalescence of etch pits and not by the progress of an etching front parallel to a single uniform lattice plane. The final shape of the edges is thus determined by the density of defects and the interplay between device geometry and acid bath hydrodynamics, as also visible in the borders of figure 2 and 3 of Ref. \cite{Plaza2021}. We note that the etching rates observed during the fabrication of this device are slightly different to those reported in Table~\ref{tab:rates}. This is expected because, in this case, the etching is just from one side, as described in Fig.~\ref{fig:proc}b, while values reported in Table~\ref{tab:rates} were calibrated in soaking condition, as in Fig.~\ref{fig:proc}c. Improved reproducibility can be obtained by calibrating the etching time for each specific hard mask geometry and etching process. 

We note that the fabrication yield of the reported devices is quite variable. While for LSMO trampolines having diagonal longer than 200\,{\textmu}m the yield is above 80\,\%, no sealed membrane with side longer than 80\,{\textmu}m survived. Our observations indicate that structural integrity is directly related to the absence of defects originated during the deposition process: if a defect is present in the suspended region (such as a trampoline tether or a sealed membrane), the structure will likely collapse. Because of this, the total device surface is the critical parameter to consider, as a larger area increases the probability of having a defect. This aspect could be improved, as an example by changing growth conditions or by using different deposition techniques.

\subsection{Bulk \ce{SrTiO3} trampoline}

\begin{figure}[]
  \includegraphics[width=\linewidth]{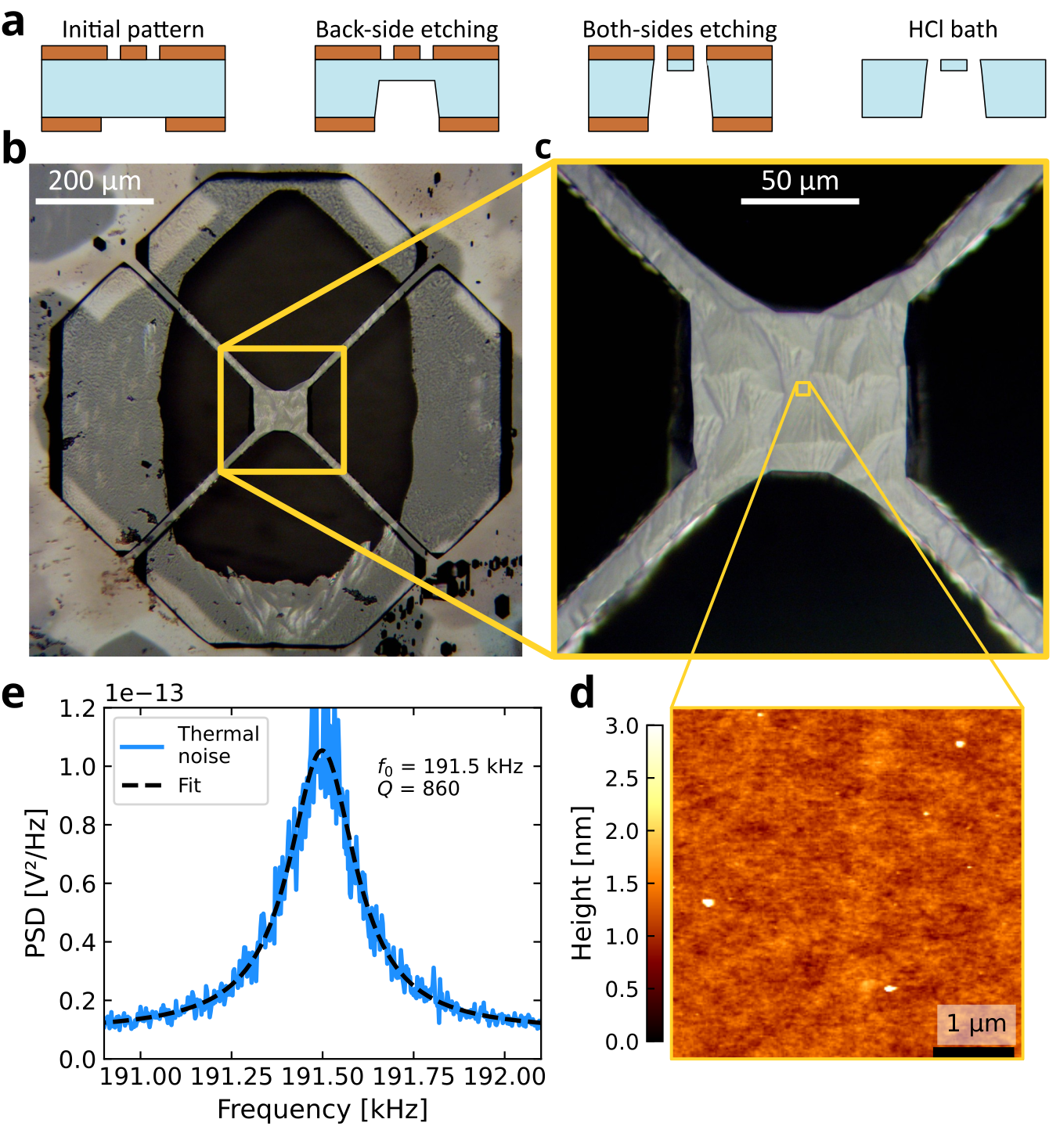}
	\caption{\label{fig:sub}
		Bulk micro-machining of a \ce{SrTiO3}(110) substrate.
		(a) Main steps of the fabrication process. 
		(b) Suspended trampoline made from the \ce{SrTiO3}(110) substrate, where the black oval region is the through-hole region.
		(c) Magnification of the pad region. The residual thickness is about 5\,\textmu{}m, so the faceting of the bottom surface is visible.
        (d) Surface topography measured by atomic force microscopy showing a RMS roughness below 0.3\,nm.
        (e) Thermal noise spectrum showing the first mechanical  mode.
              }
\end{figure}

Contrarily to the previous cases, we can fabricate suspended devices entirely made from STO employing oxides thin films deposited on both the surfaces as hard mask. This is demonstrated in Figure~\ref{fig:sub}, showing a trampoline carved from a 110\,{\textmu}m-thick STO(110) substrate. The main steps of the fabrication protocol are reported in Fig.~\ref{fig:sub}a, where, contrary to the previous cases, LSMO is employed as a hard mask on both the surfaces. The top layer is patterned as trampoline, similarly to what reported in Fig.~\ref{fig:trench}, but in this case the geometry is enlarged: the tethers are 350\,{\textmu}m-long and have a width of 35\,{\textmu}m, while the central pad is 100$\times$100\,{\textmu}m$^2$. The bottom mask is a 500$\times$750\,{\textmu}m$^2$ rectangular window. Such widening is required to let the STO structure be resilient to the under-etching during the second etching step, when the sample is soaking in the acid bath (Fig.~\ref{fig:proc}c). Since the corrosion progresses both in-plane and out-of-plane, and the corresponding rates are similar, the width of the hard mask shall be comparable to the thickness of the substrate still to be etched, i.e.\ few tens of micrometers. This guarantees that the final device will have a continuous flat surface even in the presence of defects that may locally enhance the in-plane etching rate \cite{Plaza2021}. The last fabrication step is peculiar to this process and consists in soaking the sample in 4.5\,\% HCl in water solution. This dissolves the LSMO masks, exposing the surface of the STO substrate \cite{Pellegrino2009}.  

Fig.~\ref{fig:sub}b shows the final device as seen from the top. The thickness of the suspended structure is about 5\,\textmu{}m. The black oval region is the through-hole across the STO substrate, with at its center the pad of the bulk STO trampoline. The residual frame shows areas having different colors, from white to gray. This is due to the microscope light reflecting from facets in the backside mask, where small defects triggered unintended etching processes. The width of this aperture is smaller than the length of the tethers. This mismatch can be tuned by controlling the duration of the first etching step, where a longer corrosion produces longer tethers. However, since the tethers get narrower over time, longer etching times may hinder the structural integrity or the surface quality. Because of this, the realization of desired device geometries requires precise design and alignment of both top and bottom hard masks. Fig. \ref{fig:sub}c is a magnification of the central region of the trampoline, where it is possible to observe the effect of etching anisotropy to the sides of the formerly square pad: while the vertical edges are still straight, the horizontal ones are rounded towards the tethers' clamping points. On the same picture, it is also possible to see crystal faceting on the bottom surface of the pad at about 5\,{\textmu}m below the top surface. This is due to the long etching process required to fabricate the trampoline. The quality of the top surface of the bulk STO trampoline was inspected by atomic force microscopy. The surface morphology of a 5$\times$5\,\textmu{}m$^2$ region at the center of the suspended pad is reported in Fig.~\ref{fig:sub}d. It shows a smooth surface with RMS roughness below 0.3\,nm and low density of residual cluster defects. Here, the presence of sharp terraces is not expected because the substrate was not terminated and had no thermal treatments. 

This suspended trampoline can be employed to realize new types of oxide-based MEMS. It is thus relevant to characterize its basic mechanical properties. Fig.~\ref{fig:sub}e shows the noise power spectral density around the frequency corresponding to the first mechanical mode of the STO trampoline measured using a fiber interferometer. The measured data are reported as a blue line, while the black dashed line is the best fit of the function describing the thermal noise power spectrum of a mechanical resonator \cite{Hauer2013}. The extracted quality factor ($Q$-factor) is of about 850, which is rather small if compared to other complex oxide systems \cite{Manca2022, Manca2023, Manca2024}. One reason for this is the absence of tensile strain \cite{Sementilli2022}, as the trampoline is made of the very material of its supporting frame. Another possible reason is the geometry of the tethers that have slightly different lengths and bulky clamping points with a wedge shape, where the trampoline smoothly fades into the substrate. Because of this, the resonator is not well-isolated from its frame, likely increasing the mechanical energy losses. These aspects could be partially improved by a careful design of the clamping points and backside aperture geometries.

The realization of bulk STO trampolines provides the opportunity to grow oxide thin films on top of \ce{SrTiO3} substrates having the same crystal quality as commercial ones, but with much higher aspect ratio. Also, these devices are also compatible with surface-reconstruction processes, such as annealing or termination in buffered-HF. This opens the possibility to design new experiments based, as an example, on the interplay between oxide interfacial systems and mechanical characterizations, thermal transport, or strain measurements.

\section{Conclusions}

Bulk micromachining of STO substrates allowed the realization of suspended devices with physical access from their top and bottom sides. This was achieved by employing epitaxial oxide having squared or rectangular windows as hard masks for the STO substrates. Calibration of the wet etching rate of \ce{SrTiO3}(001) and (110) substrates in controlled conditions was instrumental to the design of hard masks geometry and to predict the characteristics of the final apertures. To show the potential of this approach in realizing different kinds of full-oxide devices, we discussed three examples, all realized by employing LSMO as hard mask and device layer. This fabrication protocol can be readily extended to other device geometries and compounds.

\section{Experimental Section}
\paragraph*{UV lithography:} Patterning of the LSMO hard masks is performed by UV lithography followed by Ar ions dry etching. SPR-220-4.5 photo-resist is spin-coated at 6000\,RPM for 45\,s and then baked at 120\,$^{\circ}$C for 150\,s.
Dry etching time is about 45\,min, with an ion energy of 500\,eV and a current density of 0.2\,mA/cm$^2$. Photo-resist residues are removed by ultrasonic baths in acetone and then ethanol at room temperature.
\paragraph*{Etching rates measurements:} The geometric parameters of the aperture that were measured are depth, top width and bottom width at different times. In-plane measurements are obtained from optical micrographs, taking the initial LSMO mask as reference for scale calibration. Depth is measured by adjusting the focal distance between the top and bottom plane at each time, with an absolute error of $\sim$1\,{\textmu}m. A schematic illustration of the measurement procedure is reported in the Supporting Information Sec.~II. Depth uncertainty for the STO(001) case is estimated as $\pm$5\,{\textmu m} because of the faceting of the (001) plane. In-plane measurements error is evaluated as $\pm$5\,{\textmu m} from the roughness of the etched edges. Reported profiles are obtained as line plots connecting these measured points in a $xy$ plane.
\paragraph*{Mechanical Measurements:} Backside mechanical measurements of the LSMO trampoline resonator were acquired in a custom setup based on the optical lever detection scheme.
It includes a 670\,nm laser, a polarizing beam splitter (PBS), a quarte-wave plate ($\lambda$/4), a $\times$10 Long Working Distance objective lens (LWD), and a custom-made four-quadrant photo-diode.
Background pressure was $\mathrm{2\cdot10^{-5}}$\,mbar and temperature was kept at the constant value of 25\,{\textcelsius}.
Vector network analyzer (VNA) is a HP Agilent 4395A.
Mechanical excitation was provided by a piezoelectric element glued nearby the sample and AC-biased by the RF output of the VNA.
Mechanical measurements of the bulk \ce{SrTiO3(110)} trampoline were conducted using a fiber interferometer operating at 1550 nm at a pressure below 10$^{-4}$ mbar and at room temperature. These data were acquired by measuring the noise spectrum of the interferometric signal, showing thermo-mechanical noise of the resonator at its eigenfrequency.

\section*{Acknowledgements}

We thank Emilio Bellingeri for support in XRD measurements and Alberto Martinelli for useful discussion on the chemistry of the etching process.
This work was carried out under the OXiNEMS project (\href{www.oxinems.eu}{www.oxinems.eu}).
This project has received funding from the European Union’s Horizon 2020 research and innovation programme under Grant Agreement No. 828784.
This work was carried out within the framework of the project "RAISE - Robotics and AI for Socio-economic Empowerment” and has been supported by European Union - NextGenerationEU.
We acknowledge financial support by MUR under the National Recovery and Resilience Plan (NRRP), Project ``Network 4 Energy Sustainable Transition -- NEST'' (PE0000021).

\section*{Open Data}

The numerical data shown in figures of the manuscript 
and the Supporting Information
can be downloaded from the Zenodo online repository:
\href{http://dx.doi.org/10.5281/zenodo.13898651}{http://dx.doi.org/10.5281/zenodo.13898651}

\bibliography{Library.bib}


\newpage\newpage

\foreach \x in {1,...,8}
{
 	\clearpage
 	\includepdf[pages={\x}]{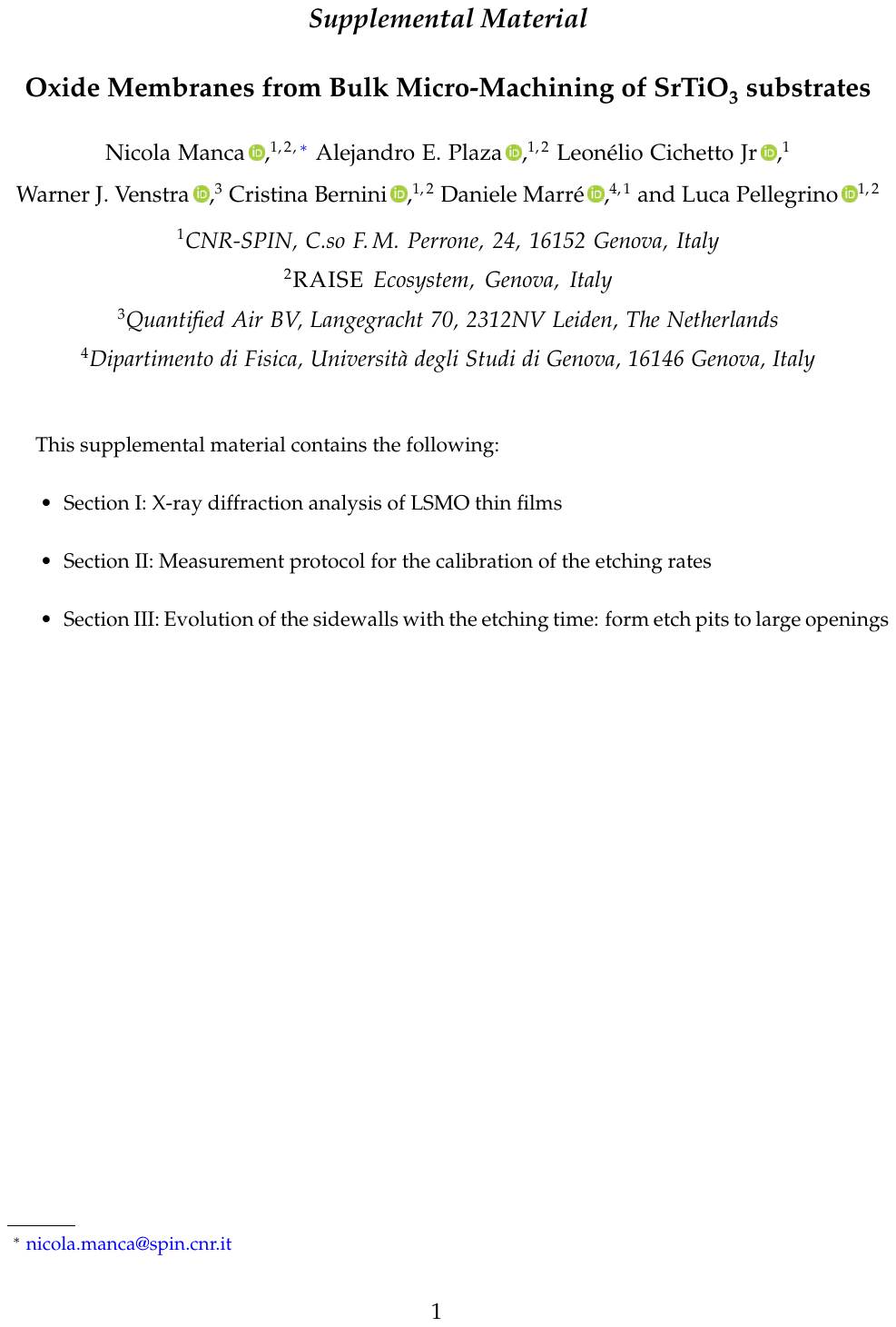}
 }

\end{document}